\documentclass[fleqn,12pt,twoside]{article}
\usepackage{espcrc1}
\usepackage{epsfig}

\title{Effects of Meson Mass Reduction on the 
Properties of Neutron Star Matter}       

\author{C. H. Hyun, M. H. Kim and S. W. Hong
\address{BK21 Physics Research Division and Institute of Basic Science,\\ 
Sungkyunkwan University, Suwon 440-746, Korea}}

\begin{document}

\maketitle

\begin{abstract}
We investigate the effects of meson-mass reduction on
the properties of the neutron star matter. 
We adopt the Brown-Rho scaling law to take into account 
density dependence of meson masses in the
quantum hadrodynamics, quark-meson coupling and 
modified quark-meson coupling models.
It is found that the equation of state becomes stiff 
when the mass of meson is reduced in dense medium.
We discuss its implication on the properties of 
the neutron star.
\end{abstract}
\vskip 0.5cm

Several theoretical studies \cite{brprl91,hlprc92,sttprc97}
based on different approaches predicted the mass 
of hadrons in medium.
Indeed, recent experimental results from
CERES/NA45 \cite{ceres} and KEK-PS E325 
\cite{kek} suggested possible reduction of 
$\rho$ and $\omega$ meson masses in medium.
In view of these previous theoretical and 
experimental studies,
inclusion of the effects of the hadron mass reduction
in dense nuclear matter is an interesting subject.

There are several models for describing nuclear matter,
some of which include quantum hadrodynamics (QHD) \cite{qhd,sw}, 
quark-meson coupling (QMC) \cite{qmc}
and modified quark-meson coupling (MQMC) \cite{mqmc} models.
In these models, coupling constants 
and some other parameters are adjusted to reproduce 
the binding energy (= 16.0 MeV) at the saturation density 
($\rho_0 = 0.17$ fm$^{-3}$).
As a means of incorporating the hadron mass change in medium,
we may employ the scaling law advocated by Brown and Rho (BR)
\cite{brprl91}
\begin{eqnarray}
\frac{m^*_\sigma}{m_\sigma}
= \frac{m^*_\omega}{m_\omega}
= \frac{m^*_\rho}{m_\rho}
= \left( 1 + y \frac{\rho}{\rho_0}\right)^{-1}.
\end{eqnarray}
In the present work we use different models for nuclear matter
and study the effects of BR scaling on
the equation of state (EOS) and particle fraction
of the neutron star matter. 
The models we use and test against each other include
QHD, QMC, MQMC, QHD with BR scaling (QHD-BR), 
QMC with BR scaling (QMC-BR) and MQMC with BR scaling
(MQMC-BR).

Neutron star matter is characterized by the charge 
neutrality ($\rho_p = \rho_e + \rho_\mu$)
and $\beta-$equilibrium ($\mu_n = \mu_p + \mu_e + \mu_\mu$)
where $\mu_i$ is the chemical potential of the 
particle $i$ ({\it i = n, p, e}, $\mu$).
These two relations determine
the number of neutrons, protons, electrons and 
muons.
\footnote{The number of muons are determined 
from the chemical equilibration condition
with electrons, $\mu_e = \mu_\mu$.}

Figure 1 shows the energy density and the pressure
with respect to the matter density.
The energy densities obtained by including the BR 
scaling increase faster than those without 
including the effects of meson mass reduction.
A similar behavior is observed in the pressure.
Fast increase of energy density and pressure 
leads to stiff EOS.
Stiff EOS of BR-scaled models is due to 
a large repulsive contribution 
generated by the $\omega$-exchanges 
in the intermediate and high densities.

Matter composition is presented in Fig. 2.
Matter composition profiles for QHD, QMC and MQMC
without including BR scaling show similar behaviors.
But when BR scaling is incorporated,
the fraction of the proton is significantly 
enhanced. 
The enhancement of proton fraction
originates from the relatively small effective
mass of the nucleon in the BR-scaled model.
\begin{figure}[t]
\begin{center}
\epsfig{file=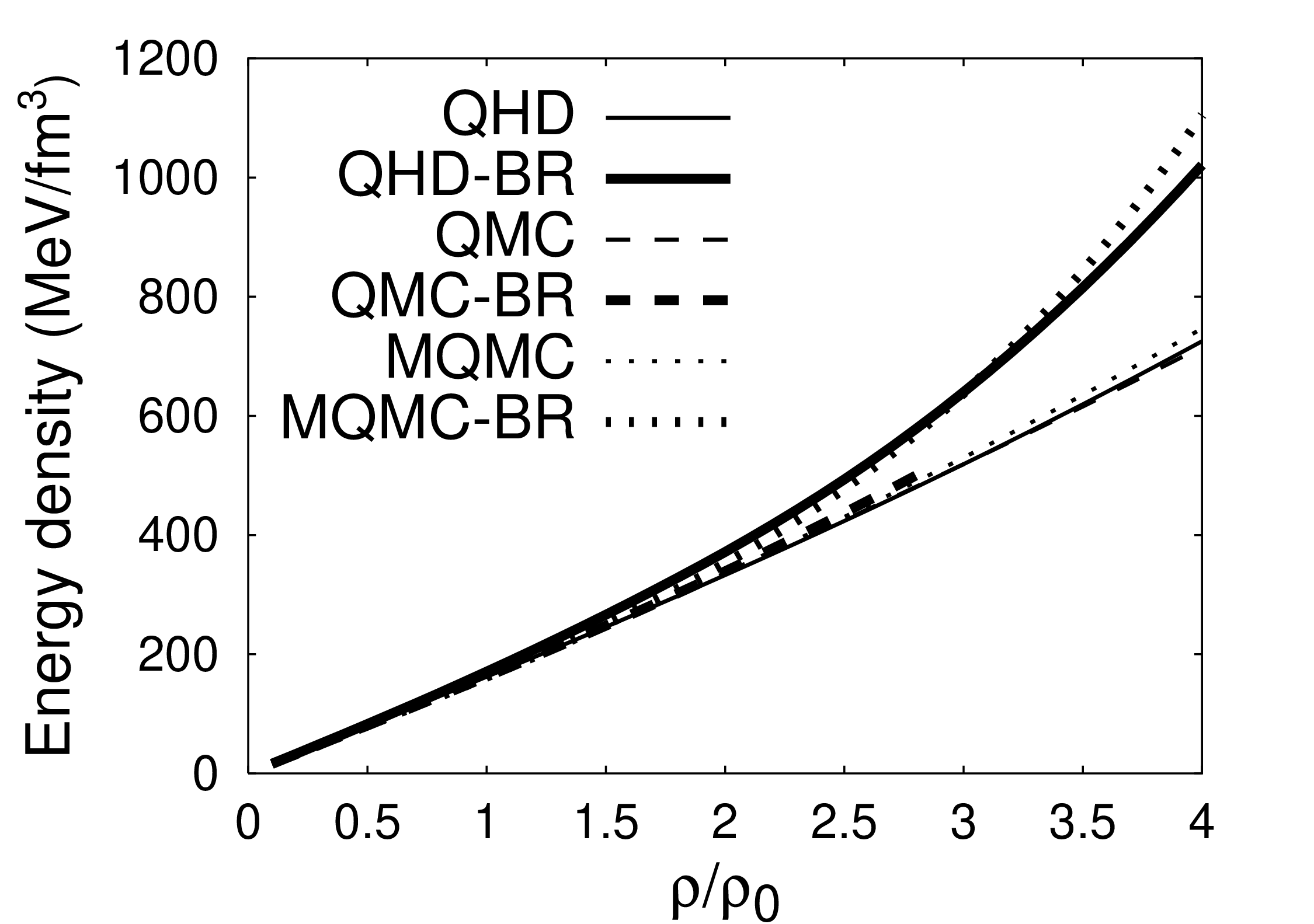, width = 5.5cm, angle = 270}
\epsfig{file=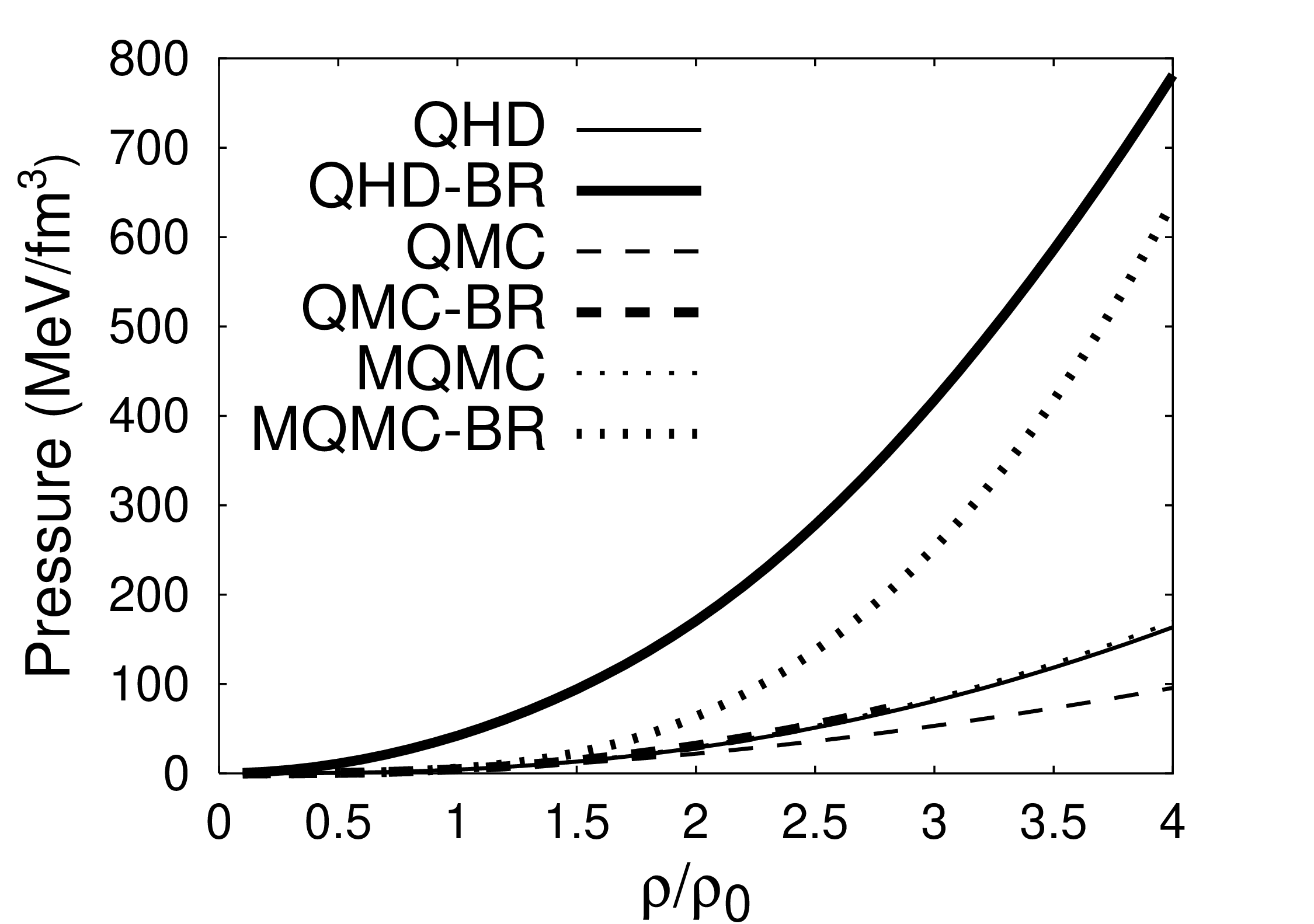, width = 5.5cm, angle = 270}
\caption{Energy densities and pressures as functions of 
the nuclear matter densities. Results from QHD, QMC and MQMC
are plotted by the thin curves. The results with including
BR scaling are plotted by the thick curves.}
\end{center}
\end{figure}

\begin{figure}[tbp]
\begin{center}
\epsfig{file=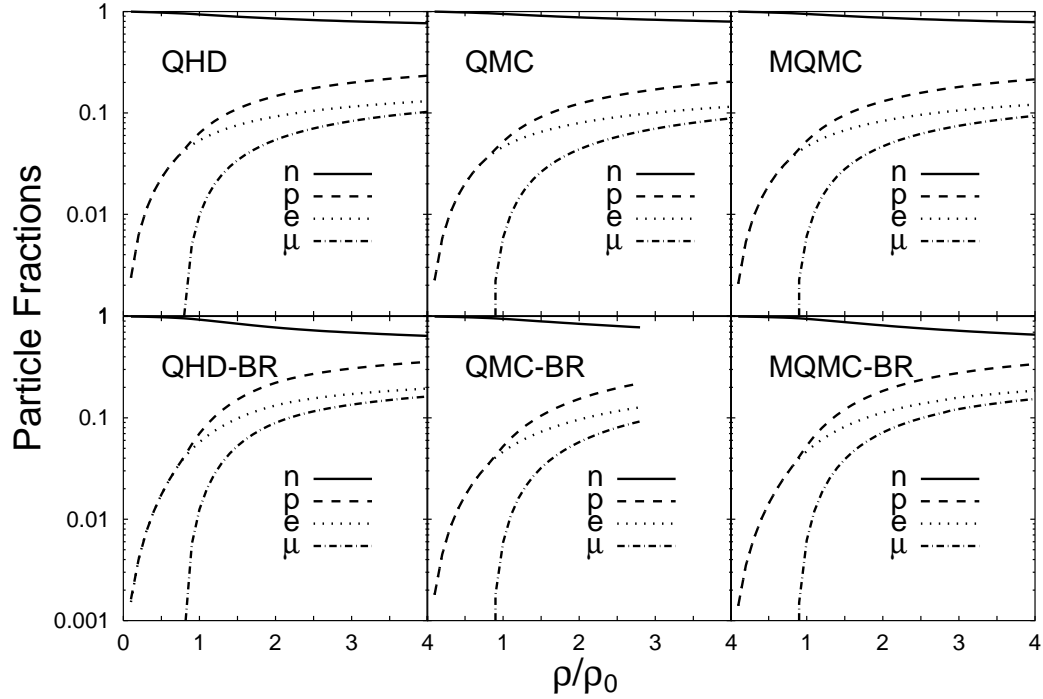, width = 10cm, angle=270}
\end{center}
\vskip -2cm

\caption{Particle ($n$, $p$, $e$, $\mu$) fractions
calculated by different models are plotted against
neutron star matter densities.}
\end{figure}


Maximum mass of the neutron star is sensitive to 
the EOS. The maximum mass of the neutron star in 
the QHD is $\sim 1.98 M_\odot$ \cite{hm98}.
As the EOS gets stiffer, the maximum mass of the 
neutron star becomes larger. 
Thus, we may expect QHD-BR to give us a
larger maximum mass.
Similarly, QMC-BR and MQMC-BR
may yield a larger neutron star mass than
QMC and MQMC, respectively.
However, more quantitative calculations require
complete inclusion of the components of the baryon octet.
When heavy baryons are taken into account,
fast-growing chemical potential of the neutron will
allow early onset of hyperon fractions, which will make 
the EOS very soft. 
Therefore the reduction of meson masses will result in 
very soft EOS which will, in turn, lower the maximum mass of 
the neutron star substantially.

Also, a large proton fraction at relatively low densities
may allow the direct URCA process 
($n \rightarrow p + e^- + \bar{\nu}_e$) to occur at low 
densities. 
Direct URCA process is known as the fastest cooling 
mechanism in the thermal evolution of the neutron star.
In addition, existence of hyperons at low densities
would cause similar fast cooling mechanism to occur efficiently
at low densities. 

In summary, we have investigated the effects of 
the meson-mass reduction on the properties of the 
neutron star matter. 
We have used the BR-scaling law as a way of incorporating
the mass reduction.
We find that the mass reduction makes the EOS stiff
and causes the proton fraction to build up quickly
at low densities,
which can lead to substantially different properties
and evolution of the neutron star.

\vskip 0.3cm
This work was supported by KOSEF (2000-6111-01-2
and 2000-2-11100-004-4).

\end{document}